# The need for single-mode fiber-fed spectrographs

March 2018

*A white paper submitted to the National Academy of Sciences Exoplanet Science Strategy*

Lead Author:

| | |
|---|---|
| Jonathan Crass | Department of Physics, University of Notre Dame, Notre Dame, IN, 46556, USA |
| | +1 574-631-2712 |
| | j.crass@nd.edu |

Co-authors:

| | |
|---|---|
| Andrew Bechter | Department of Physics, University of Notre Dame, Notre Dame, IN 46556, USA |
| Eric Bechter | Department of Physics, University of Notre Dame, Notre Dame, IN 46556, USA |
| Charles Beichman | NASA Exoplanet Science Institute, California Institute of Technology, 1200 East California Blvd, Pasadena, CA 91125, USA |
| | Jet Propulsion Laboratory, California Institute of Technology, 4800 Oak Grove Drive, Pasadena, CA 91109, USA |
| Cullen Blake | Department of Physics & Astronomy, University of Pennsylvania, Philadelphia, PA 19104, USA |
| David Coutts | Department of Physics and Astronomy, Macquarie University, Sydney, Australia |
| Tobias Feger | Department of Physics and Astronomy, Macquarie University, Sydney, Australia |
| Sam Halverson | Department of Physics & Astronomy, University of Pennsylvania, Philadelphia, PA 19104, USA |
| | NASA Sagan Fellow |
| Robert J. Harris | Zentrum für Astronomie der Universität Heidelberg, Landessternwarte Königstuhl, Königstuhl 12, 69117 Heidelberg, Germany |
| Nemanja Jovanovic | California Institute of Technology, 1200 East California Blvd, Pasadena, CA 91125, USA |
| Peter Plavchan | Department of Physics and Astronomy, George Mason University, Fairfax, VA 22030, USA |
| Christian Schwab | Department of Physics and Astronomy, Macquarie University, Sydney, Australia |
| | Australian Astronomical Observatory, Australia |
| Gautam Vasisht | Jet Propulsion Laboratory, California Institute of Technology, 4800 Oak Grove Drive, Pasadena, CA 91109, USA |
| James K. Wallace | Jet Propulsion Laboratory, California Institute of Technology, 4800 Oak Grove Drive, Pasadena, CA 91109, USA |
| Ji Wang | California Institute of Technology, 1200 East California Blvd, Pasadena, CA, 91125, USA |

MOTIVATION

Precise Doppler radial-velocity (RV) instruments will continue to play an essential role in advancing our holistic understanding of exoplanetary systems. The combination of orbital parameters from transit surveys and follow-up RV measurements is vital to unlock mass and density estimates of detected planets, giving us insight into their environment and structure. However, the exoplanet field is reaching a critical juncture: the measurement sensitivity of existing radial-velocity instruments is becoming the limiting factor in further increasing our knowledge. Without improvement in delivered RV measurement precision, we will not be able to provide dynamical mass and density estimates for some of the most exciting (and consequently most challenging) discoveries expected from new transit missions including the Transiting Exoplanet Survey Satellite (TESS) (Plavchan et al. 2015, Ricker et al. 2014). RV precisions at the 10cm/s level are required to fully confirm earth-like analogues, provide masses and measure density to the 1-5% level from these missions. Additionally, this RV capability will also be important to allow for efficient target selection for facilities such as the James Webb Space Telescope (JWST).

While stability is vital for achieving RV precision, instrument systematics are only one of the drivers behind current limitations. Perhaps one of the most daunting challenges is how to address the effects of stellar activity. This term encompasses the effects of many different stellar phenomena (including photospheric velocities, star spots, faculae and plage), and each process can introduce errors at the 1m/s level over a variety of timescales (Cegla et al. 2013, Oshagh et al. 2017). Currently, it is commonplace for observations affected by stellar activity to be identified by known indicator lines, such as the Ca II H and K lines, and there is no way to correct for the stellar imprecision imposed on these measurements. Data obtained at these times is of limited use when searching for sub-m/s RV amplitude signatures and may exclude some of the most important RV observations, specifically those at times of maximum RV amplitude. It is therefore prudent to develop instruments and techniques that can utilize these times of activity, overcoming observing inefficiencies and increasing the likelihood of detections.

Simulations indicate that asymmetries in spectral line profiles resulting from stellar activity can be measured in high resolution spectra (R>150,000) (Davis et al. 2017). If asymmetries can be measured, they can be removed in the data reduction process. The challenge with this technique is that to achieve an acceptable signal-to-noise at these high resolutions, a large collecting area and an efficient end-to-end instrument is required. Seeing-limited spectrographs require large instrument volumes to achieve this high spectral resolution due to their intrinsic properties. They are also expensive to build, difficult to stabilize (Strassmeier et al., 2008), and are challenging to design for the upcoming ELTs. Therefore another solution must be found. A promising way forward is to use single-mode fibers to inject light to a spectrograph. This mitigates many of the error terms facing current generation seeing-limited RV instruments while simultaneously offering the capability of high resolution spectroscopy within a small optical footprint (Schwab et al. 2012, Crepp 2014, Jovanovic et al. 2016a).

SINGLE-MODE FIBERS AND PRECISION RV

One of the key contributors to radial velocity systematic errors are changes in the illumination profile of the spectrograph (Spronck et al. 2015). State-of-the-art seeing limited instruments typically use multi-mode fibers (MMFs) to transport light from the telescope to a remotely-located spectrograph for stability reasons. These MMFs however suffer from modal noise and incomplete image scrambling, even after correction via agitation and other methods. Single-mode fibers overcome this effect entirely: irrespective of the illumination of the fiber, the spatial output is always a fixed Gaussian beam intensity profile.

While SMFs offer distinct advantages, they also pose a key challenge for use within astronomy: how do you efficiently couple light from a large telescope into tiny optical fibers? This has, until recently, been challenging to accomplish with the only option being to use small aperture telescopes to minimize the overall effects of atmospheric turbulence on the incident beam as has been adopted by MINERVA-RED (Blake et al. 2015) and RHEA (Feger et al. 2016). However, with the technological advancement and maturity of facility adaptive optics (AO) systems and their ability to deliver routine near-infrared (NIR) diffraction-limited beams, several groups are now developing systems to efficiently couple infrared beams into SMFs (Bechter et al. 2015, Bechter et al. 2016, Jovanovic et al. 2014, Jovanovic et al. 2016b, Jovanovic et al. 2017). Coupling efficiencies ranging from 25-47% have been demonstrated on-sky at the Subaru and Large Binocular Telescopes. With expected performance improvements through enhanced incident beam stabilization, overall AO system performance and beam reshaping technologies, efficiencies approaching their theoretical limit are expected in the next few years. These systems are ideal for feeding to an astronomical spectrograph over a wide range of seeing conditions, airmasses and in broadband light. They have the potential to deliver telescope-to-spectrograph throughputs that are comparable to seeing-limited instruments when accounting for the losses experienced during MMF scrambling.

PRECISION RVS IN THE INFRARED

Previous exoplanet studies have predominantly focused on early- and mid-type stars as they are well suited to current visible wavelength RV instruments (Fischer et al. 2016). Observations of later type K- and M-stars, with their peak emission at wavelengths approaching 1 micron, offer advantages over studies of earlier type stars: they are abundant in number, have long stellar lifetimes and offer an increased RV amplitude for the same mass of planet in the (closer-in) habitable zone. However, to study planets around late-type stars requires a high-precision infrared (IR) instrument, necessitating a thermally stabilized cryogenic environment and costly and potentially noisy IR detectors (Mahadevan et al. 2014, Hearty et al. 2014, Stefansson et al. 2016, Gao et al. 2016, Gagne et al. 2016). This has previously been beyond either the state-of-the-art technical capabilities (or project budgets) leaving these intrinsic benefits untapped.

Today, with the advancement of infrared detector technologies and the ability to couple NIR light into SMFs using AO efficiently, the exoplanet community is beginning to develop instruments specifically for this near-infrared bandpass. There is a growing demand to study later-type stars and at the Third Workshop on Extremely Precise Radial Velocities, numerous instruments covering the NIR/IR (whether under design, development or commissioning) were presented: SPIROU, CARMENES, iSHELL, NIRPS, HPF, IRD, GIARPS, iLocater, PARVI, RHEA and MINERVA-RED (Wright & Robertson 2017). These will benefit from infrared technology advances in recent years which have facilitated the ability to study exoplanets around previously inaccessible late-type stellar populations. In particular, the final four listed instruments offer additional unique capabilities – the use of single-mode fibers (SMFs) to inject light into the instrument spectrograph.

HIGH SPECTRAL RESOLUTION WITH A SMALL INSTRUMENT FOOTPRINT

The design of SMF-fed spectrographs is driven by different factors compared to their seeing-limited counterparts. As the size of the illumination input approaches the diffraction limit at the instrument operating wavelength (for example a SMF operating at a wavelength of 1 micron has an effective size of around 6 microns), the system becomes insensitive to the diameter of the telescope which feeds it (Jovanovic 2016a). Compared to the large input aperture required for seeing-limited instruments, which in turn drives the need for large optical systems (made worse when requiring high spectral resolution), SMF-fed instruments can be compact and simultaneously achieve high resolution. To maintain image quality at the diffraction limit, an SMF-fed system requires precision optics that deliver high wavefront quality, a requirement not present in the current generation of instruments. However, the optical surfaces are small, and the spectrograph camera is optically slow as the spectrograph input must be magnified onto the detector to achieve appropriate sampling (Crepp et al. 2016, Schwab et al. 2016). This makes the optics well within current fabrication capabilities and relatively inexpensive. The small beam diameter and slow camera focal ratio relaxes alignment tolerances for a given design, adding to the inherent stability of SMF spectrograph designs. Additionally, the ability to have a compact instrument design allows a cost-effective way to manufacture instruments from intrinsically stable materials such as Invar, Zerodur or Silicon Carbide.

## PRECISION RVS WITH A SMF-FED SPECTROGRAPH

While SMFs immediately remove the errors induced by modal noise, they also facilitate the mitigation of several other systematic effects. Instrument thermal stability at the milli-Kelvin level or better is required for any precision RV instrument to reduce optical changes within the spectrograph to suitable levels (e.g. grove density changes on diffraction gratings). This type of stability is simpler to achieve on a small volume compared to a seeing-limited instrument for a large aperture telescope, particularly if aided by the use of intrinsically stable materials (see previous section). It would be immensely challenging to develop a seeing-limited instrument with the high spectral resolution (R>150,000) needed to measure and correct stellar activity effects while also achieving sub-mK thermal stability. With SMFs, this becomes both achievable and affordable.

An additional source of error for seeing-limited instruments is sky-background due to the large angle subtended by their fibers on the sky. For an SMF-fed spectrograph which uses AO, the angular-size on sky is at the diffraction limit of the telescope, providing several orders of magnitude suppression in sky-background compared to their seeing-limited counterparts and reducing it to negligible levels.

## SINGLE-MODE FIBERS AND POLARIZATION

SMFs provide an ideal solution for eliminating modal noise by propagating only a single spatial mode. However, for each allowed spatial mode there exists two orthogonal polarization modes (s- and p- modes) resulting from the vector wave equation (Halverson et al. 2015). Thus a standard `single-mode' fiber is, strictly speaking, bi-modal. Mechanical stresses such as bending or twisting of the fiber during an observation cause the net polarization state exiting the fiber to evolve in time. This behavior of SMFs presents a problem for RV spectrographs as polarization changes manifest as spectral intensity modulations when interacting with polarization sensitive diffraction gratings. Uncontrolled changes in the state of polarization result in systematic radial velocity errors, the magnitude of which scales with the degree of polarization. This means that strongly polarized sources (e.g. calibrators including laser frequency combs or etalons) are more greatly impacted by these effects as opposed to stars which are typically very weakly polarized. It is expected this effect can be reduced to acceptable levels either in the data reduction process (continuum normalization) or by effectively scrambling the polarization state to a determined average over the course of an exposure (much like scrambling for multi-mode fibers). This is an active area of research and as the first generation of precision SMF-fed spectrographs come online, strategies can be developed, tested and optimized.

## ADDITIONAL BENEFITS

Seeing-limited instruments are often unable to resolve individual stars within close-binaries making it challenging to undertake studies of exoplanets in these types of systems (Wright et al. 2013). The diffraction-limited input delivered from telescope AO systems provides the opportunity to obtain RV measurements on each binary component separately, allowing studies to be undertaken which are currently impossible. Additionally, in the NIR, the use of a detector with continuous readout allows direct measurements required for chromatic barycentric correction of the Earth's motion, reducing complexity and ensuring consistency between science and calibration data, which is needed to achieve <1 m/s precision.

CHALLENGES AND TECHNOLOGICAL ADVANCEMENTS NEEDED

The use of NIR detectors for precision RV work is still a relatively new area and an ongoing effort. Additionally, unlike visible light detectors, where there is a large commercial market to drive technology advancement, NIR detectors suffer from a limited market to drive investment and a small pool of suppliers. A high resolution spectrograph with a large bandpass requires a large number of detector pixels. The largest IR focal plane array commercially available today is the H4RG detector from Teledyne (4k×4k pixels). While showing very promising noise characteristics through investment from missions such as WFIRST, a compromise must still be made between resolution, pixel sampling and wavelength coverage. To facilitate high resolution with a wide bandpass, the total number of available detector pixels needs to be increased either through larger physical detector sizes, higher pixel densities or by making detectors close-buttable, while also improving or maintaining noise properties. Investing in this technology will improve observing efficiencies and measurement precisions.

The requirement to obtain high resolution spectra to correct for stellar activity forces RV instruments to work in the photon-limited regime. Therefore, to improve signal-to-noise of individual measurements, increasing end-to-end instrument efficiency is vital. A key driver of this is the performance of the AO system injecting light into the instrument SMF. Continued improvements in AO capabilities are being driven by the needs of the upcoming ELTs. The precision exoplanet field will significantly benefit from this (both in terms of RV as well as high-contrast imaging) and as such, it should strongly be supported. Additionally, by developing automated and minimal overhead AO systems such as Robo-AO, SMF-fed spectrograph could be utilized routinely on a wide-range of telescopes worldwide, offering increased cadence of observations of targets and the possibility for increasing the number of RV instruments in a cost-effective way (Baranec et al. 2014).

Seeing-limited instruments typically use replicated gratings for their dispersing elements. These are fabricated by copying a master grating into an epoxy layer applied onto an intrinsically stable substrate (typically Zerodur). While sufficient for seeing-limited instrument where imaging quality is less important, for diffraction-limited SMF-fed spectrographs, wavefront errors induced by the grating surface may be the key contributor to image degradation through the instrument. This has the potential to limit final RV precision by inducing errors when fitting a PSF. Additionally, as changes in the line spacing of the grating directly drives instrument stability requirements, the effective coefficient of thermal expansion (CTE) of the replicated grating surface and its impact on RV precision should understood. If replicated gratings are shown to limit precision, direct ruled gratings with low CTE values will be required. However, as beam diameters and gratings are small for SMF-fed systems, these are affordable and commercially available.

Spectral contamination resulting from telluric absorption lines will be present in ground-based spectrographs (both SMF-fed or seeing-limited), especially those operating in the NIR. This may be one of the largest obstacles preventing NIR instruments achieving their expected RV precision and therefore requires improved methods and strategies for mitigation. Significant work has been undertaken by many instrument teams to address this, however, it remains a major problem. The only certain way to overcome this challenge is through a space-based mission, as is being studied as part of the Earthfinder probe study.

## FUTURE DIRECTIONS

While the current capabilities of AO systems limit the use of SMF-fed instruments to the NIR, developing SMF-fed spectrographs at visible wavelengths offers several advantages: The need for a cryogenic environment is eliminated, challenges of NIR detectors are bypassed while the ability to have high spectral resolution in a compact volume is maintained. To facilitate this however, visible light AO needs to become commonplace and routine which requires improved wavefront sensing techniques and correction capabilities. This type of development is ongoing through test-bed systems which aim to reach maximal AO performance, such as SCExAO (Jovanovic 2015, Guyon 2017).

Until such a time when visible light AO is routinely possible, the use of photonic lanterns coupled to a current generation AO systems may offer an interim solution (Robertson & Bland-Hawthorn 2012, Schwab et al. 2012, Jovanovic et al. 2017). These photonic devices take a multi-mode input and output it into several single-mode beams, each of which could illuminate an SMF-fed spectrograph. This offers an avenue to deliver the benefits of using SMFs for precision RV measurements to telescopes worldwide.

## SUMMARY

Without improvement, radial-velocity instruments will limit our ability to study some of the most interesting planetary systems from future transit missions. To provide the masses and precise densities of these systems, RV precisions at the 10cm/s level are required. Without this level of precision, the ability of the exoplanet field to continue to develop a holistic picture of planetary systems will stall.

Instrument stability is only a part of the challenge; we must develop techniques to overcome the intrinsic RV signatures imposed on spectra by stellar activity to detect planetary signatures below 1m/s. Simulations show promise that this can be achieved by using high-resolution spectra (R>150,000), however this requires a large telescope aperture to achieve sufficient signal-to-noise. For seeing-limited designs, this cannot be easily achieved in a cost-effective way while also maintaining the required instrument stability. The use of single-mode fibers fed by adaptive optics offers a viable solution and provides a pathway to study some of the most exciting systems soon to be detected.

## REFERENCES

Baranec, C. et al., 2014, ApJL, 790, L8
Bechter, A. et al., 2015, Proc. SPIE, 9605, 96051U
Bechter, A. et al., 2016, Proc. SPIE, 9909, 99092X
Blake, C. et al., 2015, AAS Meeting #225
Cegla, H. et al., 2013, ApJ, 763, 95
Crepp, J.R., 2014, Science, 346, 809
Crepp, J.R. et al. 2016, Proc. SPIE, 9908, 990819
Davis, A. et al., 2017, ApJ 846, 59
Feger, T. et al., 2016, Exp Astron 42, 285
Fischer, D. et al., 2016, PASP, 128, 066001
Gagne, J. et al., 2016, ApJ, 822, 40
Gao, P. et al., 2016 PASP, 128, 4501
Guyon, O. & Males, J., 2017, arXiv:1707.00570
Halverson, S. et al., 2015, ApJL, 814, L22
Hearty, F. et al., 2014, Proc. SPIE, 9147, 914752
Jovanovic, N. et al., 2014, Proc. SPIE, 9147, 91477P
Jovanovic, N. et al, 2015, PASP, 127, 890
Jovanovic, N. et al. 2016a, PASP, 128, 121001
Jovanovic, N. et al., 2016b, Proc. SPIE, 9908, 99080R
Jovanovic, N. et al., 2017, A&A, 604, A112
Mahadevan, S. et al., 2014, Proc. SPIE, 9147, 91471G
Oshagh, M. et al., 2017, A&A, 606, A107
Plavchan, P. et al., 2015, arXiv:1503.01770
Ricker, G.R. et al., 2014, Proc. SPIE, 9143, 914320
Robertson, J.G. & Bland-Hawthorn, J., 2012, Proc. SPIE, 8446, 844623
Schwab, C. et al., 2012, Proc. IAU, 293
Schwab, C. et al., 2016, Proc. SPIE, 9912, 991274
Spronck, J.F.P. et al., 2015, PASP, 127, 1027
Stefansson, G. et al., 2016, ApJ, 833, 175
Strassmeier, K.G. et al., 2008, Proc. SPIE, 7014, 70140N
Wright, J.T. et al., 2013, ApJ, 770, 119
Wright, J.T. & Robertson, P., 2017, Res. Notes AAS, 1,